# Measurements of Optical Kerr Nonlinearity $n_2$ in Compressed Gases

Yury E. Geints[1,*], Victor O. Kompanets[1,2], Sergey V. Chekalin[1,2]

[1]*V. E. Zuev Institute of Atmospheric Optics SB RAS, Acad. Zuev Square 1, Tomsk 634055, Russia*
[2]*Institute of spectroscopy RAS, Fizicheskaya Str. 5, Troitsk, Moscow 108840, Russia*
[*]Corresponding author e-mail: ygeints@iao.ru

## Abstract

High-power optical pulses experience self-focusing when propagating in a gaseous medium due to the manifestation of the cubic (Kerr-like) nonlinearity. The magnitude of this effect depends on the Kerr nonlinearity coefficient $n_2$, which in turn may depend on the parameters of laser radiation and the propagation medium. We present experimental data on the coefficient $n_2$ for atomic Ar, molecular $N_2$ and $CO_2$ with a pressure change from 1 to 11 bar and optical pulse duration from 50 to 500 fs of propagating femtosecond near-IR laser radiation (800 nm). Importantly, all three gases under study possess close $n_2$-values in the short pulse limit (50 fs) over the entire pressure range. According to our data, for the first time, as far as we know, the Kerr nonlinearity in $CO_2$ is obtained at atmospheric pressure equal to $n_2(CO_2) = 10.0 \pm 1.1 \cdot 10^{-24}$ m$^2$/W. Meanwhile, with increasing gas pressure, effective $n_2$ also increases due to the manifestation of aberrational pulse self-focusing caused by the development of modulation instability. In addition, according to our approximate estimate, the magnitude of the inertial component in the cubic nonlinearity of molecular gases ($N_2$, $CO_2$) is substantial and increases with both pulse duration and gas pressure.

## Introduction

The optical Kerr effect causes high-power laser radiation to propagate in transparent media in a nonlinear self-focusing mode [1]. Due to the strong spatial and temporal self-modulation of the radiation as it propagates, the pulse becomes compressed in both the longitudinal (temporal) and transverse directions, resulting in what is known as *beam collapse*. To arrest this collapse, the optical radiation fragments into localized high-intensity regions that are resistant to disturbance over a relatively long section of the optical path. These regions are known as filaments. A large number of studies have been conducted on the self-focusing and filamentation of laser pulses in different media. Several reviews on this topic have also been published, including [1-6].

Laser filamentation is a unique phenomenon that leads to light-induced ionization of the medium, the generation of plasma [7, 8], anomalous broadband optical radiation – supercontinuum [9], as well as a microwave terahertz signal [10], which can be used for remote laser diagnostics of the aerosol-gas atmosphere [11] and efficient delivery of concentrated laser energy over long distances [12, 13]. Pulse filamentation is initiated by it self-focusing (SF), which is governed by some critical power $P_c$. According to the definition [14, 15], this parameter corresponds to the situation where the natural diffraction of a laser beam is neutralized by the Kerr optical self-focusing effect. The critical power depends on the nonlinear optical properties of the medium, specifically, on the cubic (Kerr) nonlinearity coefficient $n_2$ as: $P_c \propto 1/n_2$. Therefore, knowledge of this parameter is essential for understanding the self-focusing and filamentation of optical pulses.

A significant amount of literature data on the $n_2$ coefficient is systematized on an electronic resource https://refractiveindex.info/n2 [16], where, in particular, experimental values can be found for most atmospheric and noble gases such as $N_2$, $O_2$, $N_2O$, Ar, Xe, Ne. At the same time, the relevant information for another naturally occurring gas, $CO_2$, is inexplicably missing in the $n_2$ database mentioned above and, in the works, published to date. Moreover, all known data on the Kerr refractive index are given for gases at normal pressure (1 bar), except for [17], which investigated carbon dioxide and xenon in near- and supercritical states. At the same time, the range of moderate pressures of several bars, which is important for the practical implementation of schemes for anomalous nonlinear compression of high-power radiation and the generation of attosecond pulses in gas-filled capillaries under pressure [18], has remained unexplored.

The aim of this study is to collect experimental data on the value of the Kerr nonlinearity coefficient ($n_2$) for three common gases: nitrogen ($N_2$), carbon dioxide ($CO_2$) and argon (Ar). $N_2$ and $CO_2$ have a molecular structure, while Ar is a monatomic noble gas. We consider the spectral range of a Ti:Sapphire-laser, with a central wavelength of 800 nm and a femtosecond pulse duration. A range of gas pressures up to 11 bar is investigated. Since $N_2$ and $CO_2$ molecules have an asymmetric structure, their Kerr response will have an orientational component due to stimulated Raman scattering. This means that, in addition to depending on gas pressure, $n_2$ also depends on the duration of the optical pulse. Characterizing this dependence is another goal of this work. We use a simple and intuitive method to measure the distance at which a laser beam is self-focused when the energy and/or duration of the pulse changes, as well as the pressure of the gas in the optical cell. We then apply the theory of self-focusing to extract $n_2$ values from these measurements.

## Phenomenological Self-Focusing Theory

In the theory of self-focusing of optical radiation in a Kerr medium, a phenomenological relation called the Marburger formula [19] is widely known, which makes it possible to calculate the $z_{SF}$ coordinate of the transverse collapse of an optical beam having a power $P_0$ exceeding a certain threshold value $P_c$ [14], resulting in progressive self-compression in a nonlinear medium. In functional form, the Marburger formula is written as follows:

$$z_{SF} \propto \frac{L_R^*}{\eta^k}, \qquad (1)$$

where $\eta = P_0/P_c$ is reduced pulse power, $L_R^*$ is the characteristic spatial scale, which depends on optical beam radius ($R$), spatial intensity (fluence) shape (beam quality $M^2$) and laser wavelength ($\lambda$) as: $L_R^* \propto R^2/(M^2\lambda)$, and $k$ is some exponent. In the canonical form of Eq. (1), the value $k = 0.5$ is used, which corresponds to the collapsing of an ideal unimodal beam (for example, a Gaussian beam) into a single axial filament. Note that in [20] it was shown that for realistic high-power beams characterized by a noticeable noise component in the transverse intensity profile, the exponent in (1) usually exceeds the theoretical value of ½ ($k > 0.5$) if the relative pulse power is sufficiently high, $\eta \gg 1$. This corresponds to the fact that the transverse compression of the beam becomes aberrational, leading to the formation of not a single but multiple nonlinear focus at once, which subsequently evolve into several optical filaments [21].

The nonlinear Kerr response of the medium to the optical field $E$ is usually given by the nonlinear addition $n_2$ to the linear refractive index $n_0$: $n(E) = n_0 + n_2|E|^2$. This makes it possible to introduce a critical power parameter to describe the self-focusing of optical radiation [19]:

$$P_c = 3.79\lambda^2 / (8\pi n_0 n_2), \qquad (2)$$

which is used in Eq. (1). In the simplest case, at a constant temperature and without taking into account the influence of pressure on the interaction cross-sections and spectral widths of the molecular lines of gases (subcritical pressure range), the model of direct proportionality of the optical parameters of a gas to its density, or pressure $p$ [22, 23] is valid. By introducing the relative pressure parameter, $\mu = p/p_0$, where $p_0$ is some reference (normal) pressure, we obtain an obvious barometric ratio for the nonlinear refractive index:

$$n_2(\mu) = \mu n_2(\mu = 1) \equiv \mu n_2^0. \qquad (3)$$

Then, taking into account (2) and (3), the Marburger equation (1) with explicit separation of dependence on pulse power and gas pressure will take the following form:

$$z_{SF}(P_0, p) = A \frac{L_R^*}{[\eta_0 \mu]^k}, \qquad (4)$$

where $\eta_0 = P_0/P_c(\mu=1)$, and all the parameters that are insignificant for further analysis are accumulated in the coefficient $A$.

The cubic nonlinear response of medium to the optical field contains two components having different time scales [24]. One of them is related to the polarizability of the atom/molecule due to the deformation of the electron cloud along the optical field polarization. Usually, this "electronic" component of the Kerr effect manifests itself within a few femtoseconds or less and is therefore considered almost instantaneous on the scale of an optical cycle. The second component of the Kerr response arises due to a change in the spatial orientation and shape [25] of an asymmetric gas molecule in the optical field resulting in the nonlinear polarizability of the medium which acquires an oscillating "phonon" component. These oscillations are realized at the frequencies of rotational Raman scattering of the molecule and usually are characterized by a much longer response time of the order of hundreds of femtoseconds [26].

Thus, in SF theory, the effective Kerr response of the medium is usually represented by the sum of two terms:

$$n_2^0(t) = n_K^0 + n_R^0(t). \qquad (5)$$

Next, for the convenience of analysis we will use the reduced form of this expression:

$$n_2^0(t) = n_K^0 \left[1 + f_R h_R(t)\right]. \qquad (6)$$

Here, $f_R$ accounts for the amplitude value of the relative fraction of electronic (instantaneous) Kerr $n_K^0$ and Raman (inertial) $n_R^0$ cubic responses of medium, and $h_R$ sets the time profile of the phonon component.

To describe $n_R(t)$ temporal dynamics, an approximation of a damped oscillator (or several independent oscillators [27]) is commonly used that are characterized by a Lorentz spectral contour centered at the phonon oscillation frequency $\Omega_R$:

$$h_R(t) \propto \int_{-\infty}^{t} \frac{\left(1/\tau_1^2 + \Omega_R^2\right)}{\Omega_R} \exp\left[-(t-t')/\tau_1\right] \sin\left[\Omega_R(t-t')\right] |E(t')|^2 dt' = \int_{-\infty}^{t} \Re(t-t') |E(t')|^2 dt' \qquad (7)$$

For atmospheric air consisting mainly of diatomic nitrogen molecules, the parameters of the Raman frequency and damping are [Couairon2007]: $\Omega_R$ = 16 THz, $\tau_1$ = 70 fs. As can be seen, expression (7) is a convolution of the square of the modulus (intensity) of the optical field $|E(t)|^2$ and the function of the Raman response of molecule $\Re(t)$, which leads to a non-monotonic dependence of $h_R$ within the duration of the laser pulse.

For example, Figure 1 shows the time dependence of the inertial Raman component of Kerr effect $h_R$ calculated using the model in Eq. (7) together with the function of the molecular response $\Re(t-t')$ of atmospheric air for optical pulses of different time durations $\tau_p$. As seen, in the case $\tau_p \simeq 1/\Omega_R$, a rather pronounced time delay is observed in the Raman response relative to the center of the laser pulse, which leads to an increase in the positive pulse self-phase modulation and the appearance of additional long-wavelength spectral components in pulse spectrum during the self-focusing [28]. With the inverse relation between the pulse duration and the frequency of the Raman oscillation, $\tau_p \gg 1/\Omega_R$, the time delay of the Raman response becomes insignificant (the response is adiabatic) but the response amplitude itself increases.

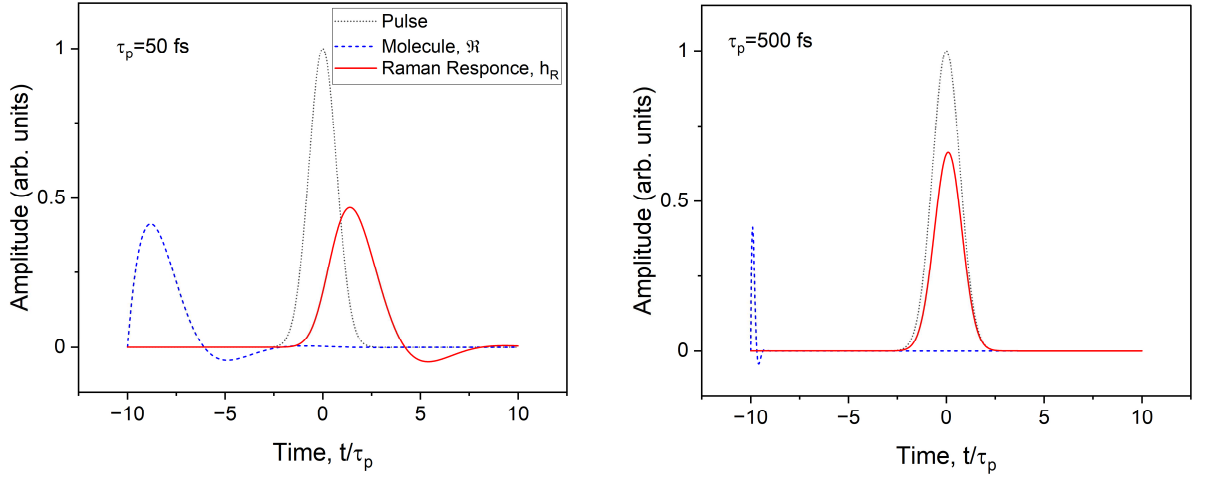

Fig. 1. A model of a damped oscillator for the inertial component of the Kerr effect $h_R$ at different laser pulse durations $\tau_p$. The time-shifted Raman response function $\Re(t-t')$ of atmospheric air is also shown.

In a real experiment, as a rule, it is not possible to trace the temporal intra-pulse dynamics of the Raman response on a sub-picosecond scale. Therefore, in this case, it makes sense to talk only about the amplitude $n_R^0$ of this nonlinear response depending on the duration of the optical pulse. In general, as follows from the model calculations in Fig. 1, it can be assumed that there are two conditional limits of "short pulse" (SP) and "long pulse" (LP) within which the amplitude of the inertial component of the Kerr effect changes from some low $n_R^{sh}$ to some high $n_R^{lng}$ values [25]. Then, Eq. (6) can be written in the following functional form:

$$n_2^0(\tau_p) = n_2^{sh}\left[1 + \Delta n_R(\tau_p)\right]. \tag{8}$$

where $n_2^{sh} = n_K^0 + n_R^{sh}$ represents the effective Kerr response in SP limit, $\Delta n_R(\tau_p) = f_R(\tau_p)(n_R^{lng} - n_R^{sh})/n_2^{sh}$ denotes the relative Raman response component, and $f_R(\tau_p)$ is some function, changing from zero to unity, which takes into account the temporal inertia of the Raman response.

Returning to the relation for the SF range (4) and taking into account the time dependence (8), within the framework of the linear barometric model (3) we obtain:

$$z_{SF}(P_0, p) = AL_R^*\left[\eta_0 \mu (1 + \Delta n_R)\right]^{-k}, \tag{9}$$

Here, the parameter of the reduced optical pulse power $\eta_0$ is defined in terms of the minimum value $n_2^{sh}$ of the nonlinear refractive index in the SP mode.

This expression indicates a simple way to quantify the temporal dynamics of changes in the inertial component of the Kerr response of the medium $\Delta n_R$. To this end, the SF distance $z_{SF}$ should be measured with a several change in the pulse duration $\tau_p$ and compared with similar measurements for a pulse having a minimum temporal length (SP mode) but with a proportionally lowered initial power $P_0$ (pulse energy $W_0$). Then, writing down the ratio of the measured $z_{SF}$ values at equal initial reduced power $\eta_0$ and considering that $\Delta n_R\left(\tau_p^0\right) = 0$, we obtain the obvious ratio:

$$\Delta n_R\left(\tau_p\right) = \left[\frac{z_{SF}\left(\tau_p\right)}{z_{SF}\left(\tau_p^0, \eta_0\right)}\right]^{-1/k} - 1 \equiv \left[\frac{z_{SF}^{lng}}{z_{SF}^{sh}}\right]^{-1/k} - 1. \qquad (10)$$

Recall that here the parameter $\eta_0$ in SP regime should correspond to the actual value of the reduced power in the LP mode, i.e. $\eta_0 = W_0\left(\tau_p^0\right)/\tau_p P_c$.

It is important to note that Eq. (10) is obtained under the condition of a constant (fixed) pressure in the gas medium. In the general case, the Marburger-like Eq. (9) has an inversely power-law dependence on the relative density of the gas $\mu$. In addition, there may be a pressure dependence of the instantaneous Kerr additive $n_2^{sh} = n_2^{sh}(p)$. Therefore, pulse SF length can become an arbitrary (unknown) function of the gas pressure even if the pulse power is fixed. A quantitative assessment of this effect can be made by correlating the $z_{SF}$ values at different pressures $p$ in the SP limit. Then, according to Eq. (9) we obtain:

$$\frac{n_2^{sh}(p_2)}{n_2^{sh}(p_1)} = \frac{\mu_1}{\mu_2}\left[\frac{z_{SF}^{sh}(p_1)}{z_{SF}^{sh}(p_2)}\right]^{1/k} \rightarrow \frac{n_2^{sh}(p)}{n_2^{sh}(p_0)} = \frac{1}{\mu}\left[\frac{z_{SF}^{sh}(p_0)}{z_{SF}^{sh}(p)}\right]^{1/k}. \qquad (11)$$

Here is assumed that $p_1 = p_0$.

In fact, the appearance of any pressure dependence on the left side of relation (11) will indicate the inclusion of new physical mechanisms unaccounted for in Eq. (4) in the process of pure cubic self-focusing dynamics of a pulse with increasing medium density. The most likely here are the manifestation of the aberrations in the transverse intensity profile and self-steepening of leading pulse edge [29] during temporary compression due to increased dispersion effects in a dense gas. Indeed, an increase in gas pressure $p$ at a fixed initial laser pulse power is formally equivalent to an increase in the reduced power $\eta$ due to a decrease in the critical power $P_c$. This markedly provokes aberrational focusing of the pulse due to a sharp increase in intensity on the axis [30] and the development of modulation instability of the radiation amplitude due to the inevitable presence of a noise component [31]. Obviously, a denser gas is characterized by a shorter group velocity dispersion (GVD) length $L_{GVD} \propto 1/\mu$ [22]. Moreover, since according to Eq. (4), $z_{SF}(\mu) \propto 1/\mu^k$, where $k < 1$, a specific situation may arise at a certain pressure when the GVD length $L_{GVD}$ is close to the length of the Kerr collapse: $L_{GVD}(\mu) \simeq z_{SF}(\mu)$. Therefore, in this case, the dispersion effects will actively compete with the SF process and distort the time profile of the pulse.

## Experimental Method

The experiments are carried out using the Spitfire Pro XP femtosecond laser system (Spectra Physics), powered by a titanium-sapphire crystal and generating femtosecond pulses with a repetition rate of up to 1 kHz at a central wavelength of $\lambda_0 = 800$ nm with a half-wavelength pulse duration $\tau_p = 45$ fs. In the experiments, the pulse repetition rate is lowered to 20 Hz for preventing the influence of a thermal lens and possible gas ionization products created in the nonlinear focus on the optical filament position.

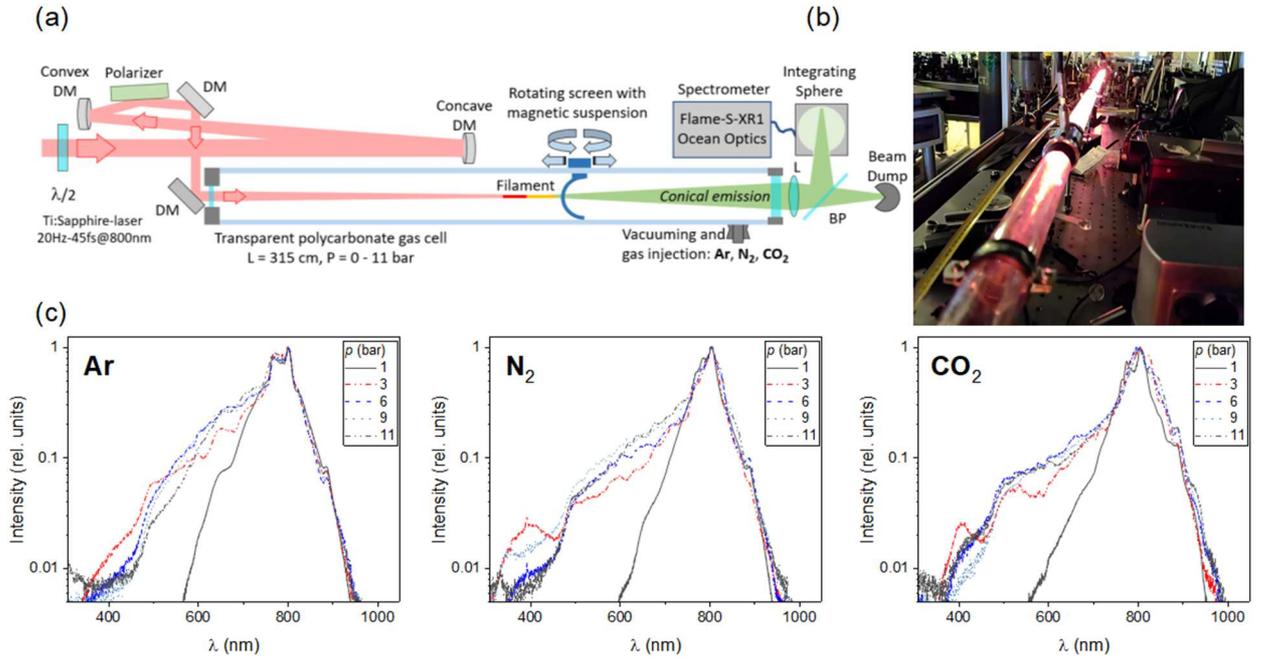

Fig. 2. (a) Experimental scheme (DM dielectric mirrors, BP fused silica plate at Brewster angle). (b) a photograph of 315 cm polycarbonate optical cell with a laser filament inside. (c) Typical spectra of the filament in various gases at the outlet of the cell, depending on the gas pressure.

Femtosecond optical radiation is focused using convex and concave mirrors to a diameter of 4 mm and then directed into an optical cell containing pressurized gas. The optical cell is a thick-walled polycarbonate tube with a diameter of 4 cm and length of 315 cm, which can withstand gas pressure up to 15 bar. The entrance and exit windows of the cell are made from fused silica, with the entrance window measuring 2 mm in thickness and the output window 6 mm in thickness.

The maximum energy of the input pulse in the experiment is 3 mJ, which corresponds to a peak power of approximately 60 GW. The pulse energy was reduced by means of an optical frequency-preserving polarizing mirror (OAFP, Avesta Project Ltd.), which reflects at a sliding angle, by rotating the polarization using an achromatic half-wave plate mounted on a wide beam towards the telescope. To extend the duration, the pulse is chirped using an internal compressor in a regenerative amplifier. Gas pressure is measured with a calibrated vacuum gauge with accuracy class 1.6, ranging from 0 to 16 bar. Gases such as argon (Ar), nitrogen ($N_2$), and carbon dioxide ($CO_2$) with high purity and a volume fraction of at least 99.995% are used at pressures from 0 to 11 bar to fill the optical cell. Argon is a monatomic gas, while the other two are molecular gases, which results in increased optical activity due to additional rotational degrees of freedom in the molecules.

For instance, Fig. 2c illustrates the spectra of filaments exiting the optical cell for three different gases under varying pressures. As can be observed, all gases exhibit a substantial broadening of the femtosecond pulse spectrum, resulting in the formation of a supercontinuum (SC), as previously reported in our work [32]. However, our focus in present study is not on the spectral characteristics of the filament but rather on its spatial location when altering the optical pulse parameters and the gaseous environment.

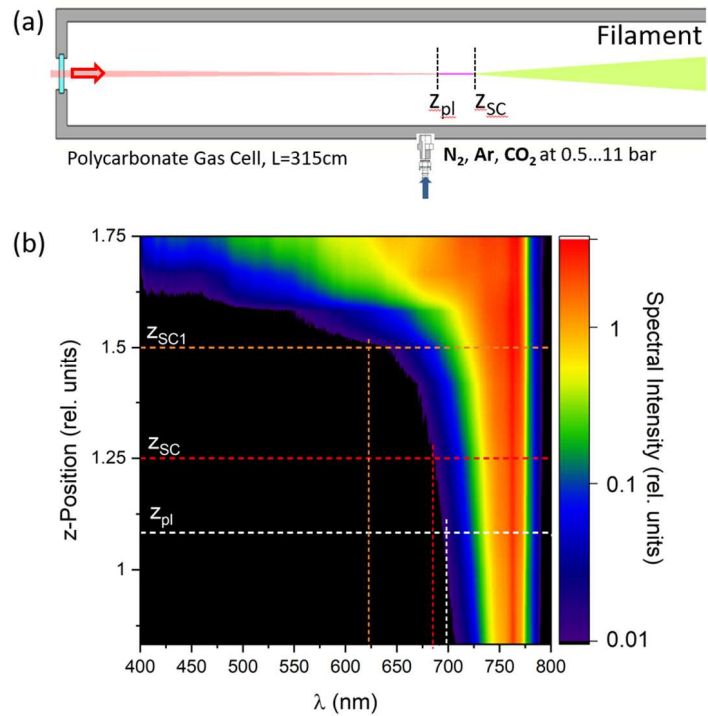

Fig. 3. The method for calculating the SF-distance of a pulse. (a) Schematic $z$-position of the measured $z_{pl}$, $z_{SC}$ and $z_{SC1}$ coordinates. (b) Pulse spectra reflected from the blocking screen in argon (at 5 bar) when the beam stopper is moved.

The method for obtaining the cubic nonlinear optical coefficient $n_2$ of the gases under investigation is schematically shown in Fig. 3a. It involves determining the location of the nonlinear focal point $z_{SF}$ of a femtosecond laser beam undergoing Kerr self-focusing as its duration, energy, or gas pressure change. To this end, a semicircular metal screen is placed inside a transparent polycarbonate tube, blocking the optical beam's path at a specific location. The screen is attached to the cuvette wall with a magnet located outside the cuvette and can be moved along the tube. To measure the radiation spectrum at the end of the cell, the screen can rotate around its axis, and due to its semicircular shape, it does not interfere with the pulse propagation.

The spatial position of the start of optical pulse filamentation is defined as the $z_{SF}$ point, which can be visually detected by a red halo appearing on the screen around a bright spot on the laser beam axis. In order to improve the accuracy and reproducibility of measurements, the $z_{SF}$ position is calculated as an arithmetic mean of two coordinates: $z_{SF} = \frac{1}{2}(z_{pl} + z_{SC})$. The first coordinate, $z_{pl}$, corresponds to the appearance of a bright white spot in the center of the beam, which indicates the beginning of the plasma region. The second coordinate, $z_{SC}$, corresponds to a red glow visible on the periphery of the beam (image on a movable screen), respectively. This position corresponds to the point where supercontinuum generation begins.

At long pulse durations, the red glow may not be clearly visible, so a third coordinate, $z_{SC1}$, is measured when the halo around the center of the beam is already bright but has not yet turned orange. This position can be considered the coordinate of beam exit from the nonlinear focus, and averaging is performed of three positions in this case.

All the spatial positions for calculating the SF distance are determined based on the spectra of the filament reflected laterally from the curtain directly through the walls of the transparent tube. These positions are also visually monitored by observing the color change in the reflected radiation. A shortpass filter with a cutoff wavelength of 780 nm is used for spectral measurements. An example of such a spectrum near the beginning of filamentation in argon at a pressure of 5 bar is shown in Fig. 3b. It can be seen that the spatial positions of $z_{pl}$ and $z_{SC}$ correspond to the cutoff wavelengths (at the 1/100 level) of 700 nm and 680 nm, respectively. The appearance of a bright red color in the filament spectrum at 630 nm indicates the position of the additional $z_{SC1}$ measurement point.

### Results and Discussion

Refer to the results of measurements of the SF length of femtosecond pulses in compressed gases. Typical dependences of $z_{SF}$ values on the gas pressure $p$ in the cell are shown in Fig. 4(a-d) using the example of carbon dioxide ($CO_2$) and pulses of different energies. Note, that graphs with the same peak power $P_0$ are grouped on each panel, but obtained in a different way. The black symbols display $z_{SF}$ when the pulse energy $W_0$ changes, but the duration remains the same, $\tau_p = \tau_p^0 = const$, i.e., in SP mode. The red symbols show the SF length at constant energy $W_0 = const$, but varying due to pulse chirping to the duration of $\tau_p > \tau_p^0$. Worthwhile noting, the initial laser pulse with a duration of 45 fs after passing through the input quartz window of the cell is extended to 48 fs. Therefore, when processing experimental data, for simplicity, the minimum pulse duration is set to $\tau_p^0 = 50$ fs.

Additionally, analytical dependencies are constructed on two panels (a) and (d) by approximating experimental data using a function as in Eq. (9):

$$z_{SF}^{fit}(\mu, \eta) = L_R^* \left[ \eta \mu \cdot a (1 + b\mu) \right]^{-0.5}, \quad (12)$$

where the characteristic diffraction scale of the beam $L_R^* = 2.68$ m is determined from experiment, and $a$ and $b$ are the fitting parameters. In this case, parameter $a$ represents the amplitude of the Kerr response $n_2$, and the parameter $b$ takes into account its change with gas pressure. Note, that in (12), $\mu = p / 1\,\text{bar}$, and $\mu = P_0 / 60\,\text{GW}$.

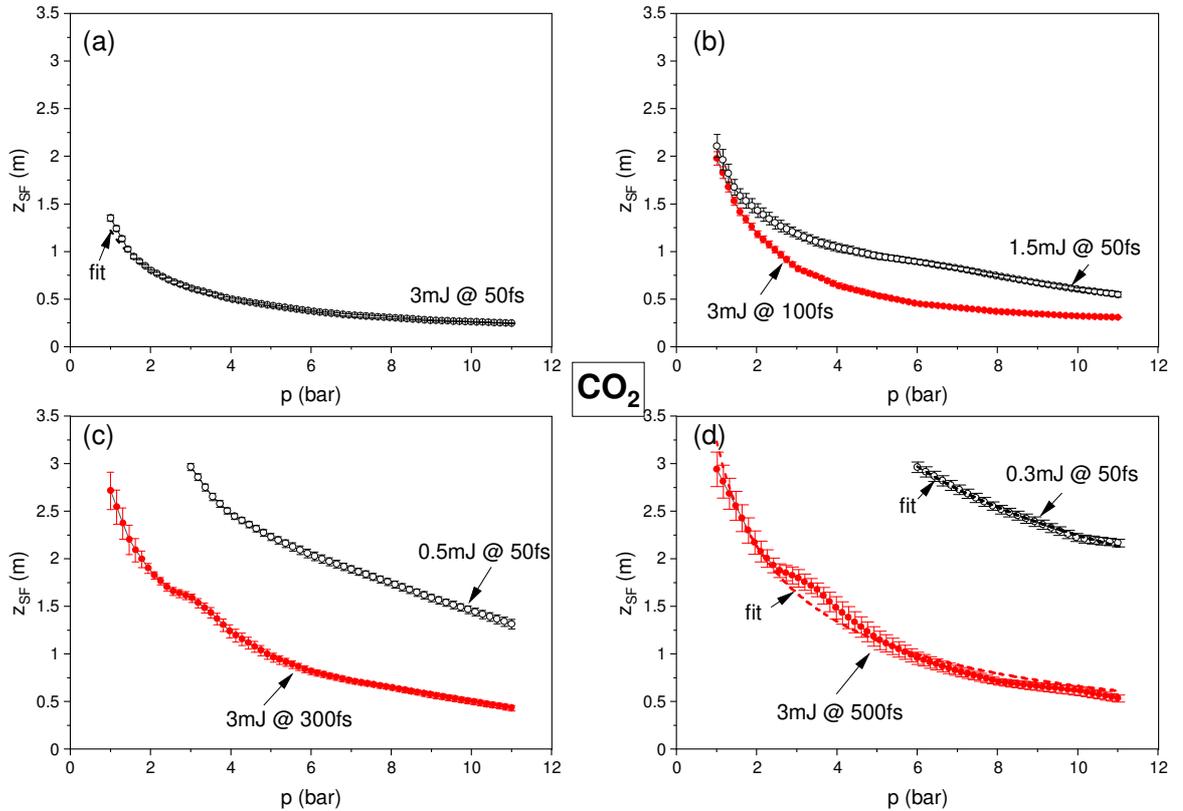

Fig. 4. The SF length $z_{SF}$ of femtosecond radiation in compressed $CO_2$ at peak optical pulse power (a) 60 GW, (b) 30 GW, (c) 10 GW, and (d) 6 GW, obtained either by changing the energy $W_0$ in the SP mode (black symbols) or by pulse chirping to the duration $\tau_p$ (red symbols). The dotted curves show the regression according to (12).

Based on the examination of these data, it can be inferred that as the gas pressure in the cell rises, the distance over which the pulse travels before forming a clearly visible filament is expected to decrease. The relationship between the SF distance and the gas pressure can be accurately approximated by Eq. (1) with the exponent $k$ equal to 0.5, which follows the well-known Marburger formula upon considering the obvious dependence: $\eta \propto 1/P_c \propto n_2 \propto \mu$. Furthermore, the standard errors of the regression are minimal for the data obtained in the SP regime and do not exceed 10%.

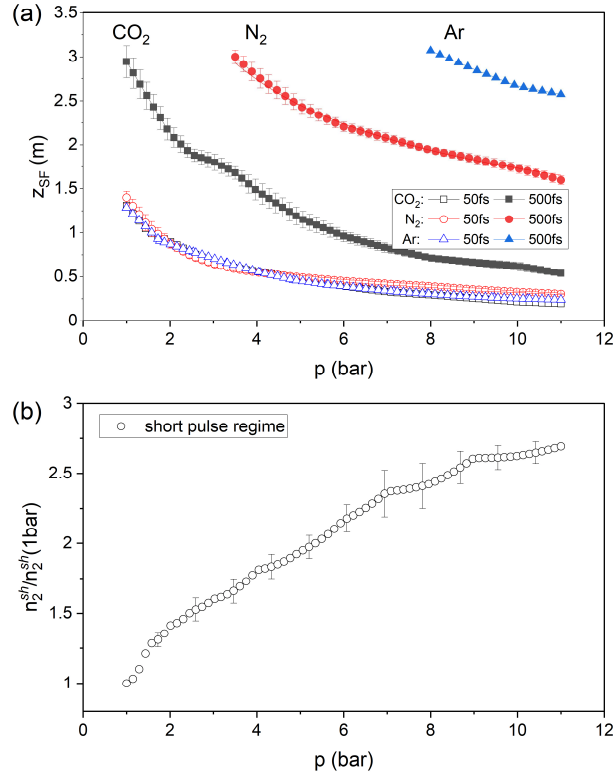

Fig. 5. (a) Pressure dependence of the SF distance of optical pulse ($W_0 = 3$ mJ) in various gases at the duration of the chirped pulse $\tau_p = 50$ fs (SP mode) and 500 fs (LP mode). (b) Kerr nonlinear refractive index $n_2^{sh}$ of $CO_2$ as a function of pressure.

Moreover, the graphs clearly demonstrate the difference in the SF distance of the femtosecond pulse in the SP and LP modes, indicating a substantial contribution of the inertial Raman component to the total cubic optical nonlinearity of the gas. This difference begins to be seen already for $\tau_p = 100$ fs (Fig. 4b) and persists even at normal $CO_2$ pressure. As the laser pulse duration increases and the gas pressure rises, the difference in SF distances increases as a whole. With a maximum duration of $\tau_p = 500$ fs (Fig. 4d), i.e. at the lowest possible peak power, the ratio $z_{SF}^{sh}/z_{SF}^{lng}$ reaches four times at $p = 11$ bar. It is worth noting that the absence of data on the SF distance in Fig. 4c and 4d indicates that at such power levels, filamentation within the three-meter cuvette was not observed in the SP mode, although the diameter of the laser beam at the output window could noticeably decrease. Furthermore, for a longer chirped pulse, a substantial Raman contribution of $\Delta n_R$ to the optical nonlinearity triggered the onset of filamentation within the cuvette containing carbon dioxide, even at $\tau_p = 500$ fs and $p = 1$ bar.

Interestingly, for pulse durations of 300 fs and 500 fs (Figs. 4c and 4d) the SF distance shows a small "hump" in the pressure range from approximately 2.5 to 4 bar. For shorter pulses, such a hump is not observed. We attribute this deviation from the inverse square-root Marburger's dependence (red dashed curve) to a change in the laser filamentation mode from single to multiple.

We have obtained similar data on the SF distance in other gases. For example, Fig. 5a compares the considered pressure dependences for $CO_2$, $N_2$, and Ar in the regimes of "short" and "long" pulses having the same energy, $W_0 = 3$ mJ.

Importantly, in the SP regime, all three gases are characterized by similar values of Kerr nonlinearity $n_2^{sh}$, despite the fact that two of them consist of rather heavy molecules, and one (Ar) is atomic. According to the literature data on self-focusing of optical pulses with a wavelength of 800 nm and a duration of $\tau_p = 40$ fs [16, 25], nitrogen and argon are known to have similar nonlinear refractive coefficients: $n_2(\tau_p) \equiv n_2^{sh} = 7.4 \cdot 10^{-24}$ m$^2$/W for $N_2$ and $9.7 \cdot 10^{-24}$ m$^2$/W for Ar. This actually gives a similar Kerr response in magnitude and should lead to similar $z_{SF}$ values as can be seen from Fig. 5.

For the carbon dioxide gas in the specified spectral range and pulse duration, we could not find any published literature on $n_2$. There is only one experimental study [17] that provides a value for the nonlinear Kerr additive of $n_2 = 3.4 \pm 0.6 \cdot 10^{-22}$ m$^2$/W, which was obtained by observing the filamentation of 200 fs pulses of a Cr:forsterite laser in an optical cell filled with compressed $CO_2$. This value is significantly higher than those observed in other gases. However, it is important to note that in the above cited work, the laser pulse was much longer than the spectrally limited 50 fs pulse used in our measurements for the SP mode. Additionally, the radiation wavelength was different, at 1240 nm. Therefore, it would be inappropriate to directly apply the experimental information from Ref. [17] to interpret the results in Fig. 5a. Simultaneously, based on the analysis of the $z_{SF}$ data presented in this graph, it can be inferred that in the case of the SP limit, the nonlinear response of $CO_2$ is most similar in magnitude to that of argon. Consequently, we can estimate that the value of $n_2^{sh}(CO_2) \cong 10^{-23}$ m$^2$/W best suits the experimental data presented for $\lambda = 800$ nm, $\tau_p = 50$ fs, and a pressure of 1 bar.

Using Eq. (11), we can construct the pressure dependence of the instantaneous Kerr additive in the SP mode. This dependence is illustrated in Fig. 5b, specifically for compressed $CO_2$, as the results will be similar for the other gases under study. As mentioned earlier, the emergence of this dependence suggests a departure from the dynamics predicted by Marburger's SF theory. $n_2^{sh}$ increase with gas pressure suggests that the reason for this deviation is the beam aberration during the pulse self-focusing due to the development of modulation instability as the medium density increases. In fact, the development of amplitude inhomogeneities in the laser beam results in a reduction in the spatial scale over which the entire pulse is shifted. In the context of model (9), this corresponds to an increase in the parameter $\eta_0$ with increasing gas pressure $p$, and consequently $n_2^{sh}$ increases also. It is worth mentioning that the transition from a single to multiple filamentation mode as the gas pressure increases has been previously explored in our research [32].

Recall Fig. 5a and focus on the SF distance and the relationship between $z_{SF}$ and the pulse duration. It is evident that a tenfold increase in $\tau_p$ to 500 fs results in a tenfold decrease in the peak power of the radiation. According to the theoretical prediction (1), this should lead to the moving away of the beam collapse point (nonlinear focus). This increase is observed in the figure for all gases. However, the change in $z_{SF}$ parameter varies dramatically depending on the gas in the cell. For example, the maximum shift in the beam collapse point occurs in argon, while the minimum shift is typical for carbon dioxide. Nitrogen falls in between these two extremes. Nevertheless, the nature of the pressure dependence $z_{SF}(p)$ remains unchanged and can still be well approximated by Eq. (12).

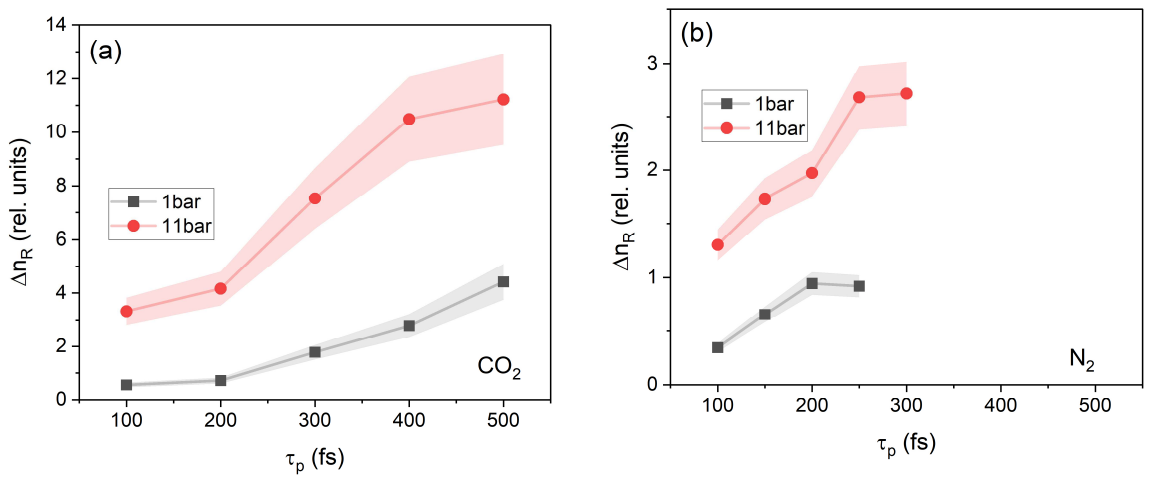

Fig. 6. The relative amplitude of the Raman nonlinearity $\Delta n_R$ (a) of carbon dioxide and (b) of nitrogen, depending on the duration of the chirped pulse at different gas pressures. The shaded areas show the standard deviation (STD) of the values.

Consider that the argon molecule, being a noble gas with no rotational degrees of freedom, exhibits a purely instantaneous Kerr response ($\Delta n_R = 0$). Therefore, we can assume that the change in the SF distance in argon is solely due to a variation in the parameter $\eta_0$, at least for a pressure of 1 bar. Consequently, the onset of filamentation in other gases compared to argon is linked to the emergence of an inertial Raman response in molecules ($\Delta n_R \neq 0$).

The relationship between $\Delta n_R$ and the length $\tau_p$ of chirped laser pulse is depicted in Figs. 6a and 6b for two molecular gases at various pressures within the optical cell. The curves in these figures were obtained by analyzing the SF distance data using the method outlined by Eq. (10). It is important to note that the relative magnitude of the Raman component is shifted along the vertical axis, with respect to the instantaneous Kerr response $n_2^{sh}$ in the SP mode.

The rise in the inertial component of the cubic response of gases as the laser pulse lengthens is evident. Furthermore, in $N_2$ and $CO_2$ at atmospheric pressure, the amplitude of the Raman additive remains comparable at the minimum chirp, with $\Delta n_R(1\text{bar}) = 0.43$ and $0.55$, respectively. However, for the longest possible pulse, the discrepancy in the values of $\Delta n_R$ becomes more pronounced: $\Delta n_R(1\text{bar}) = 0.92$ ($N_2$) and $4.4$ ($CO_2$). It is worth noting that the absence of experimental data on nitrogen at $\tau_p > 300$ fs is due to the limitation of the measurement distance by the length of the optical cell.

Moreover, Fig. 4b demonstrates that in nitrogen, the change in Raman response, $\Delta n_R$, saturates at $\tau_p = 250$ fs, regardless of the pressure. In contrast, in carbon dioxide (Fig. 6a), the saturation of the Raman response either does not occur at all (at $p = 1$ bar) or occurs for much longer pulses. This is likely due to the intricate structure of the gas molecule, which in the case of triatomic carbon dioxide is characterized by a large moment of inertia and a more pronounced spatial asymmetry, resulting in a greater number of rotational degrees of freedom with different characteristic response times [33].

Surprisingly, the contribution of the Raman effect to the cubic nonlinearity in the LP regime is quite substantial. As mentioned earlier, at standard pressure in the cell, $\Delta n_R$ is comparable in magnitude to the purely Kerr response in nitrogen and significantly surpasses it in carbon dioxide. In a previous experimental study [34], the authors reached the same conclusion when examining the Kerr nonlinearity coefficient in a number of gases, including nitrogen, xenon, and nitric oxide. At elevated pressure, the dominance of the inertial molecular response becomes even more pronounced, potentially exceeding the Kerr response by an order of magnitude for carbon dioxide.

The outcome of our investigation is presented in Table 1, which displays the values of the effective nonlinear coefficient of cubic nonlinearity $n_2$ for three gases under various pressures for two limiting regimes of SP ($\tau_p$ = 50 fs) and LP ($\tau_p$ = 250…500 fs). The experimental data from [25] were used as a benchmark for the purely Kerr response of nitrogen and argon.

Table 1. Optical Kerr nonlinearity ($n_2 \times 10^{-24}$ m$^2$/W) of compressed gases at $\lambda$ = 800 nm.

| Molecule | 1 bar | | 5 bar | | 11 bar | |
|---|---|---|---|---|---|---|
| | SP, $n_2^{sh}$ | LP, $n_2^{lng}$ | SP, $n_2^{sh}$ | LP, $n_2^{lng}$ | SP, $n_2^{sh}$ | LP, $n_2^{lng}$ |
| Ar | 9.7±1.5[a] | - | 18.9±1.8[b] | - | 26.2±2.5[b] | - |
| $N_2$ | 7.4±0.9[a] | 14.1±1.5[b] | 14.4±1.6[b] | 36.0±3.9[b] | 20.1±2.2[b] | 74.7±8.5[b] |
| $CO_2$ | 10.0±1.1[b] | 54.1±3.8[b] | 19.6±2.0[b] | 121.5±9.3[b] | 27.3±2.6[b] | 333.1±10.4[b] |

[a] Ref. [25]

[b] This work

Note that the ratio of the effective Kerr nonlinearity obtained in our work for nitrogen at normal pressure in the SP and LP modes is $n_2^{sh}/n_2^{lng} \approx 0.52$. This value is more than twice as high as the similar ratio reported in [25], where $n_2^{sh}/n_2^{lng} \approx 0.23$. This suggests that, in our measurements, the contribution of inertia to the cubic nonlinearity of nitrogen is significantly lower.

In our opinion, the reason for this discrepancy in the data may be due to a different method of measuring the cubic response of the medium used in our experiments. We directly measured the filamentation starting position of collimated femtosecond radiation using the well-established theoretical model of self-focusing. In [25], a classical pump-probe technique was employed, which measures the phase shift between a reference and test beam in a specific volume of interaction. In the latter case, the value of $n_2$, recovered from indirect measurements, significantly depends on the length of the beam interaction region. This length can change with a change in the duration (or power) of the main pulse, despite the focused geometry of the experiment. This can lead to inaccuracies in the processing of empirical information.

Furthermore, the methodology employed in our experiments is not free from errors, which, as noted above, may be attributed to the aberration of the pulse, which leads to multiple filamentation of the radiation. This makes it more challenging to accurately determine the position of the nonlinear focus of an optical pulse, and it may lead to an increase in the effective Kerr coefficient $n_2^{sh}$ derived from measurements of the SF coordinate in the SP mode. However, the data provided indicate that the relative contribution of inertial molecular nonlinearity to the total cubic nonlinearity of gaseous $N_2$ and $CO_2$ is substantial. In fact, it can be several times or even an order of magnitude greater than the instantaneous electronic response.

As a final remark, we note that employing the method of phase modulation, also known as chirping, to change the power of a laser pulse can result in a dynamic change in its duration during its propagation through a medium with frequency dispersion. This phenomenon is attributed to the presence of the dispersion in the group velocity of optical pulse and can be directly derived from the equation governing the propagation of a short wave-packet [35]. In this scenario, the sign of the phase modulation, which controls the order of the spectral components within the pulse envelope, determines whether the pulse will contract or, conversely, expand along the time axis as it traverses the so-called dispersion range: $L_{GVD} = \tau_p^2 / k_\omega''$, where $k_\omega''$ is the GVD coefficient. For most gases, the value of this parameter is relatively low, typically ranging from 0.3 to 0.6 fs$^2$/cm [16]. Consequently, even at a pressure of 10 bar, taking into account the direct proportionality $k_\omega''(p) = p k_\omega''(p=1)$ [22, 23, 32], the dispersion length of the 50-fs pulse will always be several meters, which is much longer than the distance of the Kerr self-focusing, according to the data in Fig. 2. This implies that in our experiments, the chirp sign was not a crucial factor and had no significant impact on the measurement of the $z_{SF}$ parameter. This is also supported by the results of our separate experiments (not shown here), where the $z_{SF}$ distance was measured with the same amplitude but different chirping parameters.

## Conclusion

In conclusion, we present experimental measurements of the self-focusing distance of high-power femtosecond pulses of an 800 nm titanium-sapphire laser in three different gases, namely Ar, $N_2$, and $CO_2$, placed in an optical cell at different pressures from 1 to 11 bar. The main objective of the study is to determine the magnitude and pressure dependence of a cubic nonlinear additive to the refractive index of the gases under study. All the experimental data obtained are processed based on the well-known formalism of the Kerr self-focusing theory of an optical beam in a cubic nonlinear medium. To clarify the magnitude of the contribution of the inertial (Raman) component to the cubic nonlinearity of gas medium, our studies are conducted with chirped pulses with varying time duration.

We found that, within the limits of a short, spectrally-limited radiation pulse ($\tau_p$ = 50 fs), all three gases under study show similar values for the coefficient of effective cubic nonlinearity across the entire pressure range. The self-focusing distance's dependence on gas pressure can be well approximated by a function that is similar to the Marburger formula, with an exponent of $k$ = 1/2, considering the pressure scaling of the critical power parameter.

Moreover, based on our data, we were able to estimate, for the first time, the Kerr nonlinearity coefficient for femtosecond radiation in $CO_2$ at atmospheric pressure. This coefficient is equal to $n_2(CO_2) = 10.0 \pm 1.1 \cdot 10^{-24}$ m$^2$/W. With increasing gas pressure, the effective $n_2$ value also increases, which we believe is due to aberrational self-focusing of the pulse caused by the development of modulation instability as the medium's density increases. In addition, it was found that the magnitude of the inertial component in the cubic nonlinearity $\Delta n_R$ in molecular gases ($N_2$, $CO_2$) is quite large and increases with both pulse lengthening and increasing gas pressure. Even at normal pressure in the optical cell, $\Delta n_R$ is comparable in amplitude to the purely Kerr response in $N_2$ and significantly exceeds it in $CO_2$. At an increased pressure of 11 bars, the predominance of the inertial molecular response becomes even more noticeable, and the ratio of the effective Kerr nonlinearity for long and short pulses for nitrogen and carbon dioxide gases can be approximately 4 and 12, respectively.

Expectably, the experimental data obtained on the amplitudes, time, and pressure dependences of the cubic nonlinearities of the studied gases may be useful in designing nonlinear optical modulators and converting optical radiation into a supercontinuum using gas-filled cells.

**Funding**
Russian Science Foundation (24-12-00056); Ministry of Science and Higher Education of Russian Federation (IAO SB RAS); project FFUU-2022-0004 of the Institute of Spectroscopy of the RAS

**Conflict of Interest**
The authors have no conflicts to disclose.

**Data availability**.
Data underlying the results presented in this paper may be obtained from the authors upon reasonable request.